\documentclass{ws-p8-50x6-00}

\begin{document}

\title{Nucleon Momentum Distributions From a Modified Scaling Analysis of
Inclusive Electron-Nucleus Scattering.}

\author{J. Arrington}

\address{Argonne National Laboratory, Argonne, IL, USA}


\maketitle

\abstracts{Inclusive electron scattering from nuclei at low momentum transfer
(corresponding to $x \geq 1$) and moderate $Q^2$ is dominated by
quasifree scattering from nucleons. In the impulse approximation, the cross
section can be directly connected to the nucleon momentum distribution via the
scaling function $F(y)$.  The breakdown of the $y$-scaling assumptions in
certain kinematic regions have prevented extraction
of nucleon momentum distributions from such a scaling analysis. With a slight
modification to the $y$-scaling assumptions, it is found that scaling functions can be
extracted which are consistent with the expectations for the nucleon momentum
distributions.}

Quasielastic (QE) electron scattering can provide important
information about the distribution of nucleons in nuclei.  With simple
assumptions about the reaction mechanism, functions can be deduced
that should scale ({\it i.e.} become independent of momentum transfer),
and which are directly related to the nucleon momentum distribution.
The concept of $y$-scaling of the quasielastic response was first proposed\cite{westandco}
in 1975.  It was shown that in the plane wave impulse
approximation (PWIA) a scaling function, $F(y)$, could be extracted from the
inclusive cross section which was related to the nucleon momentum distribution.
In the simplest approximation, the scaling variable $y$ is the initial
momentum of the struck nucleon along the direction of the virtual photon.
We determine $y$ from energy conservation assuming a spectator model of the
interaction and neglecting the transverse momentum of the struck nucleon:
\begin{equation}
\nu + M_A = \sqrt{M_N^2 + (y + q)^2} + \sqrt{M_{A-1}^2 + y^2},
\label{eqn.ydef}
\end{equation}
where $M_A$ is the mass of the target nucleus and $M_{A-1}$ is the ground state
mass of the $A-1$ nucleus (assumed to be in an unexcited state).

Measurements of inclusive electron-nucleus scattering from deuterium and
heavy nuclei at $x >1$ have been
performed\cite{89008data}  at JLab up to $Q^2 \approx 7$ GeV$^2$.
At low $Q^2$ values the scaling function
depends strongly on $Q^2$ due to final state interactions (FSIs).  As
these FSIs become small the extracted scaling function
becomes nearly independent of $Q^2$ and depends only on $y$, as predicted in
the $y$-scaling picture.  However, while the data show scaling in $y$,
this by itself does not ensure that the
scaling function is connected to the momentum distribution.  We present
here an attempt to test the assumptions of the scaling analysis and
the extraction of the nucleon momentum distributions.

\begin{figure}[tb]
\begin{center}
\epsfxsize=24pc 
\epsfbox{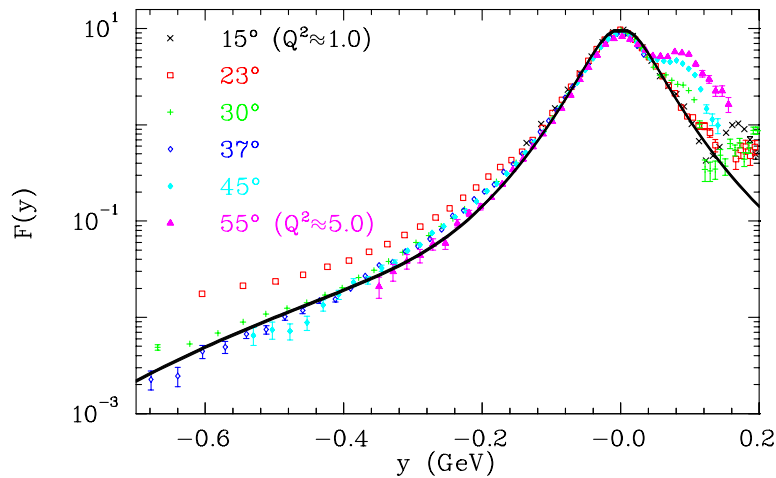} 
\caption{Scaling function $F(y)$ for deuterium from E89-008 for scattering
angles between 15 and 55 degrees, after subtracting a model of the inelastic
contributions (large for $y>0$, generally negligible for $y<0$).
Errors shown are statistical only.  The solid line represents the expected
$F(y)$ based on a calculation of the deuteron momentum distribution using
the Av14 NN potential.  \label{fig.yd}}
\end{center}
\end{figure}


Figure~\ref{fig.yd} shows $F(y)$ for deuterium, as extracted from the cross
sections measured in E89-008\cite{89008data}.  As the momentum distribution is related
to the derivative of $F(y)$, the lack
of high precision data on deuterium at large $|y|$ makes it difficult to
directly extract the momentum distribution from this data.
We can, however, compare the scaling function to what we expect based on a
calculation of the deuteron momentum distribution.
The solid line is a calculation of $F(y)$ using a momentum distribution 
calculated from the Argonne-v14 N-N potential.  
The normalization of the scaling function extracted from the data is consistent with unity (as it
must be if it is related to the momentum distribution) and the distribution is in
generally good agreement with the calculation.  In particular, they are in
very good agreement at very large values of the nucleon momentum, ($y$).
This region is especially
important because these high momentum components are generated by short range
interactions of the nucleons.  It has been suggested that
final state interactions in this region, where the nucleons are close together,
may not disappear as $Q^2$ increases.  If there were large final state
interactions that were nearly
independent of $Q^2$, one might see scaling but the scaling function would
not yield the proper momentum distribution in the tails.  The data from
deuterium indicate that such $Q^2$-independent FSIs are small or absent,
although higher precision data at high $Q^2$ and large $|y|$ would allow
a much stronger limit to be set.


While the $y$-scaling analysis of the deuterium data appears to yield the
correct deuteron momentum distribution, this is not the case for the heavier
nuclei.  The momentum distribution extracted from $F(y)$ for heavy nuclei
falls off much more rapidly at large $y$, indicating that the high momentum
components in heavy nuclei are much smaller than in deuterium, which is the
opposite of what one might expect.  In addition, the normalizations of the scaling
functions for heavy nuclei are $\sim$20-30\% lower than they should be if $F(y)$
is related to the nucleon momentum distribution.  These problems indicate that
there is a failure of some kind in the scaling analysis for heavy nuclei.
The breakdown for $A>2$ nuclei comes from the assumption that the residual $(A-1)$ nucleus
remains in an unexcited state.  This is a
reasonable approximation when removing a single nucleon from a shell
at low missing energy.  However, the high momentum nucleons are
predominantly generated by short range correlations, meaning that the momentum of the struck
nucleon is mostly balanced by a single nucleon, leaving a high momentum
nucleon in the residual nucleus.
In the following analysis, we take this into account by assuming a
simple three-body breakup of the nucleus, where the struck nucleon is assumed
to be one of a correlated pair of nucleons moving within the residual $(A-2)$
nucleus.  The scaling variable in this case is $y^* = k + K_{2N}/2$, where
$y^*$ is the total momentum of the struck nucleon, coming from the relative
momentum of the two correlated nucleons, $k$, and the momentum of the
pair within the residual nucleus, $K_{2N}$.  

\begin{figure}[tb]
\begin{center}
\epsfxsize=24pc 
\epsfbox{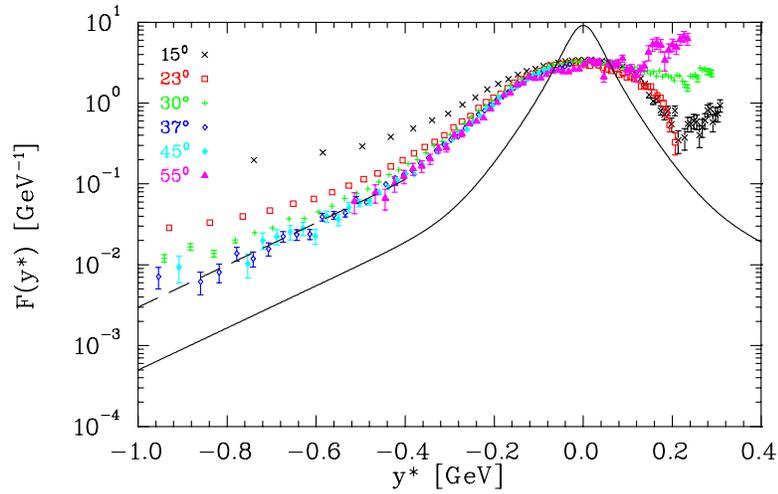} 
\caption{Scaling function $F(y^*)$ for iron from E89-008, after subtracting
a model of the inelastic contributions which dominate for $y>0$.
The solid line is a fit to the measured $F(y^*)$ for deuterium.  The dashed line is
the tail of the deuteron fit, scaled up by a factor of six.  \label{fig.yfe}}
\end{center}
\end{figure}

Figure~\ref{fig.yfe} shows $F(y^*)$ from iron, along with the fit to the
deuterium scaling function (note that for deuterium there is no $(A-2)$
residual nucleus, so $y^*=y$).  The high momentum behavior is identical for
deuterium and heavy nuclei (carbon, iron, and gold),
indicating that the two-nucleon correlations that dominate in
deuterium are the main source of high momentum components in heavy nuclei.
Using the modified scaling variable, the normalization of $F(y^*)$ is also
consistent with unity, as it should be if the scaling function is related to
the nucleon momentum distribution.

While this data indicates that the modified scaling analysis is valid and
allows extraction of the nuclear momentum distributions, the data at large
nucleon momentum is somewhat limited, especially for few-body nuclei where
the extracted distributions can be compared to essentially exact calculations
of nuclear structure.  Future measurements are planned with 6 GeV
beam\cite{e02019} which
will significantly increase the amount of data in the scaling region at large
nucleon momenta.  This data will significantly improve the
data at large $|y|$ and $Q^2$, and will include measurement on $^3$He 
and $^4$He.  We can then use this to extract information on the momentum
distributions in heavy nuclei, and study in more detail the nature of their
short range correlations.

This work is supported (in part) by the U.S. DOE, Nuclear Physics Division,
under contract W-31-109-ENG-38.

\end{document}